\documentclass[12pt]{article}
\usepackage{latexsym}
\usepackage{amsmath}
\usepackage{rotating}
\usepackage{tikz}
\renewcommand{\epsilon}{\varepsilon}

\usepackage{amsmath,bm}
\usepackage{amssymb}
\usepackage{xcolor}
\usepackage{amsthm}
\usepackage{mathtools}
\usepackage[ansinew]{inputenc}
\usepackage{graphicx}
\usepackage{epsfig}
\usepackage{dsfont}
\usepackage{bbm}
\usepackage{url}
\usepackage{placeins}
\usepackage[FIGTOPCAP]{subfigure} 
\usepackage{multirow}
\usepackage[font=footnotesize,labelfont=bf]{caption}
\usepackage{authblk}
\usepackage{longtable}

\usetikzlibrary{matrix,calc,arrows,shapes,trees}
\usetikzlibrary{shapes.geometric, arrows,positioning}
\usetikzlibrary{backgrounds}

\tikzset{
  startstop/.style={
    rectangle, 
    rounded corners,
    minimum width=3cm, 
    minimum height=1cm,
    align=left, 
    draw=black, 
    thick,
    fill=blue!30
    },
      dummy/.style={
    rectangle, 
        rounded corners,
    minimum width=3cm, 
    minimum height=1cm, 
    align=left, 
    draw=white, 
    fill=white
    },
      dummy2/.style={
    rectangle, 
        rounded corners,
    minimum width=0.05cm, 
    minimum height=1cm, 
    align=left, 
    draw=white, 
    fill=white
    },
  process/.style={
    rectangle, 
        rounded corners,
    minimum width=3cm, 
    minimum height=1cm, 
    align=left, 
    draw=black, 
    fill=blue!10
    },
  decision/.style={
    rectangle, 
    rounded corners,
    minimum width=3cm, 
    minimum height=1cm, align=center, 
    draw=black, 
    thick,
    fill=green!30
    },
  arrow/.style={draw,thick,->,>=latex},
  dec/.style={
    ellipse, 
    align=center, 
    draw=black, 
    fill=green!30
    },
}

\usepackage{natbib}
\newtheorem{satz}{Theorem}

\newtheorem{algorithm}[satz]{Algorithm}

\def\3{\ss}

\newcommand{\bea}{\begin{eqnarray*}}
	\newcommand{\eea}{\end{eqnarray*}}
\newcommand{\be}{\begin{eqnarray}}
	\newcommand{\ee}{\end{eqnarray}}

\newcommand{\ba}{\begin{array}}
	\newcommand{\ea}{\end{array}}

\def\3{\ss}

\begin{document}
	
	\title{{\bf \Large Identification of changes in gene expression 
   }}

	\author[1]{L. Ameis} 
	\author[1*]{K. M\"ollenhoff}

	\small\affil[1]{Institute of Medical Statistics and Computational Biology (IMSB), Faculty of Medicine, University of Cologne}\normalsize

	\date{}
	\pdfminorversion=4
	\maketitle
	
		\begin{abstract}
Evaluating the change in gene expression is a common goal in many research areas, such as in toxicological studies as well as in clinical trials. In practice, the analysis is often based on multiple t-tests evaluated at the observed time points. This severely limits the accuracy of determining the time points at which the gene changes in expression. Even if a parametric approach is chosen, the analysis is often restricted to identifying the onset of an effect.
In this paper, we propose a parametric method to identify the time frame where the gene expression significantly changes. This is achieved by fitting a parametric model to the time-response data and constructing a confidence band for its first derivative. The confidence band is derived by a flexible two step bootstrap approach, which can be applied to a wide variety of possible curves. Our method focuses on the first derivative, since it provides an easy to compute and reliable measure for the change in response. It is summarised in terms of a hypothesis test, such that rejecting the null hypothesis means detecting a significant change in gene expression. Furthermore, a method for calculating confidence intervals for time points of interest (e.g. the beginning and end of significant change) is developed. 
 We demonstrate the validity of our approach through a simulation study and present a variety of different applications to mouse gene expression data from a study investigating the effect of a Western diet on the progression of non-alcoholic fatty liver disease. 
		\end{abstract}
		
		
		\parindent 0cm
		
		\maketitle
  
		\vspace{0.5cm}
* Corresponding author: Kathrin M\"ollenhoff, eMail: kathrin.moellenhoff@uni-koeln.de

\section{Introduction} \label{sec0}
One common objective in the field of  RNA-sequencing, based on the measurement of transcriptomes (see \cite{WG09}), is to identify differentially expressed genes. Possible examples include dose-gene expression data (see, e.g., \cite{KK13}) or time-gene expression data (see, e.g., \cite{WDM21}). 
In recent times, there have been calls among researchers that a similar large-scale project on a par with the Human Genome Project  (see \cite{HumGen} and \cite{HumGenSite}) for RNA is required (see \cite{Pen24} and \cite{NASEM24}). It is therefore of great importance to develop new methods for the analysis of RNA sequencing data. 

\cite{AH10} introduced an approach to analysing gene count data using the negative binomial distribution, that is employed in the well-known DESeq \texttt{R} package. Its successor -- the DESeq2 packages (see \cite{LH14}) -- has been cited over $35,000$ times as of July 2024 (see \cite{Dese2Cite}). However, as described in the vignette of DESeq2 (see \cite{Deseqvig}), the analysis of the data 
includes the dose/time as a discrete covariate, which limits the precision of the analysis to the measured dose levels/time points. 

One question, where this problem is particularly evident, is the identification of an alert concentration, defined as the lowest measured concentration where the difference between the concentration and  the control significantly
exceeds a  critical level of the relevant response
 (see \cite{DF11}) or the lowest concentration with a noticeable effect (see \cite{JK19}). 
 Traditionally, the analysis is based on applying multiple $t$-tests or -- as recommended -- Dunnett tests (see \cite{Hot14}), focussing on the measured dose levels. Consequently, with these tests it is not possible to identify a dosage between those levels. 
In order to address this issue, 
\cite{KG21} proposed a model-based approach assuming a monotonic relationship between dose and response. This method allows for a continuous identification of the lowest concentration at which
the response significantly exceeds the response for the control by a given threshold. This method was further developed by \cite{MS22} through the introduction of a  procedure that is independent of the assumption of monotonicity and replaces an asymptotic calculation of the variance by a parametric bootstrap.  

When analysing time-response data, a similar problem is given by the identification of the time frame of significant changes in gene expression. For example, a change in gene expression is indicative of  deregulation and thus the presence of a biological process. In a two-group situation, a difference in the time periods can indicate the deceleration or acceleration of an underlying process.  Furthermore, a gene ontology (GO) analysis of genes that undergo a change in expression over a specified time frame allows inference regarding the processes that occur at the organism level.
In \cite{WDM21} so called rest-and-jump genes (RJG) were observed. These are genes that showed delayed deregulation, either reaching a plateau afterwards or continuing to deregulate. In this context, it is of interest to consider the entire time frame during which the gene expression changes, rather than focusing only on the beginning of an effect.

Based on this idea, in this paper we propose a model-based approach that allows the identification of the entire time frame within which the change in gene expression is considered significant. This is accomplished by formulating a hypothesis test considering a test statistic based on the absolute value of the first derivative of a selected continuous parametric model with regard to a threshold $\lambda\geq0$. 
Rejecting the null hypothesis indicates the presence of  a slope, that is significantly larger than $\lambda$ and thereby a significant change in gene expression. The test decision is based on the construction of a lower simultaneous confidence band for the first derivative, inspired by the method developed in \cite{MS22}. 
Within the period of significant change in gene expression,
it may be of interest to focus on specific time points, such as the beginning and end of the significant change. The second goal of this paper is to construct confidence intervals for these time points.

This paper is structured as follows: First, we introduce the method to estimate the time frame of significant change in gene expression, based on calculating the first derivative of the estimated time-response model and the estimation of a corresponding lower simultaneous confidence band. Based on this, the method is extended, allowing for the estimation of confidence intervals for time points of interest. Second, the methodology is subjected to a simulation study. Finally, a number of potential applications are illustrated using a data set from the western diet mice study (see \cite{WDM21}), a toxicological study that examines the impact of a western diet (WD) on the progression of non-alcoholic fatty liver disease (NAFLD). 

\section{Methodology}\label{sec1}
The fundamental idea of this method is to initially fit a model to the data and subsequently assess the change in gene expression by considering its first derivative. This is possible since the first derivative describes the slope of the gene expression curve at any time point $t$, and thus can be regarded as a measure for the change. Once a suitable model has been fitted, the next step is to develop a hypothesis test that tests if and when the first derivative significantly exceeds a chosen threshold, denoted $\lambda$. 
The corresponding time point marks a change in gene expression. 
The first three steps of the procedure are inspired by the method developed in \cite{MS22} adjusted for the first derivative, while the fourth and last step provides the possibility to obtain a confidence interval for the identified time points. 

\subsection{Model fit and derivation}\label{subsec:model}
We regard a data set with $n$ observations in total, that are conducted at $m$ different time points $t_p$, $p=1,\dots,m$. Let $n_p$ denote the number of observations at time point $t_p$, i.e. $\sum_{p=1}^mn_p=n$
. Furthermore, the study period is defined as $T=\left[t_1,t_m\right]$, while the study duration is expressed as $\overline{t}=t_m-t_1$.
The initial step is to fit a parametric model to the data. Hereby, a suitable model can be selected from a pool of possible models using a selection criteria, as, for example, the AIC (see \cite{FK22}). We define
$$y_{p,q}=f(t_p,\theta)+\epsilon_{p,q}  \ \ \ p=1,\dots,m, \ \ \ q=1,\dots,n_p,$$
where $f$ denotes the selected model. We assume that the errors  are independently and identically normally distributed $\epsilon_{p,q}\sim N(0,\sigma^2)$. The parameters $\theta\in B\subset \mathbb{R}^r$ are estimated by applying the OLS method, i.e.
$\hat{\theta}=\arg \min_{b\in B}\sum_{p=1}^m\sum_{q=1}^{n_p}\left(y_{p,q}-f(t_p,b)\right)^2$,
and, due to the assumption of normally distributed errors, coincide with the corresponding ML estimates, so that the AIC can easily be calculated. The corresponding variance estimate is given by $\hat{\sigma}^2=\frac{1}{n-r}\sum_{p=1}^m\sum_{q=1}^{n_p}\{y_{p,q}-f(t_p,\hat{\theta})\}^2.$
However, as we are interested in detecting significant changes, the method developed here does not focus on the model $f$ itself, but rather on its first derivative $f'$.
Throughout this paper, we will focus on applying the method using the 4pLL (also called sigmoid Emax) model and the beta model, respectively, as parameterized in \cite{BP09}. Nevertheless, the method is applicable to all parametric models with a first derivative that can be calculated analytically.
Exemplary, the first derivative of the 4pLL model is given by
$$f'_{4pLL}(t,(a,b,c,h))= \frac{b\cdot c^h\cdot t^{h - 1} \cdot h}{(c^h + t^h)^2}$$
    for $\theta=(a,b,c,h)\in \mathbb{R}^4$. The first derivative of the beta model is
    \begin{align*}f'_{beta}(t,(a,b,\delta_1,\delta_2))=&b \cdot B_\delta  \left(\frac{t^{\delta_1-1}\left(1-\frac{t}{scal}\right)^{\delta_2-1}\left(\delta_1 scal-(\delta_1+\delta_2)t\right)}{scal^{\delta_1+1}}\right)  \end{align*}
    for $\theta=(a,b,\delta_1,\delta_2)\in \mathbb{R}^4$, where $scal$ denotes a fixed scaling parameter and $B_\delta=\frac{(\delta_1+\delta_2)^{\delta_1+\delta_2}}{\delta_1^{\delta_1}\delta_2^{\delta_2}}$.

\subsection{Hypothesis test for the identification of significant changes}\label{subsec:test}
Here, we introduce a hypothesis test in which the rejection of $H_0$ leads to the assertion of a significant change in gene expression. In the previously introduced parametric framework this means that $f'$ exceeds a threshold $\lambda\geq0$ at some time point t, which leads to the hypotheses 
\begin{align}\label{hypotheses}H_0: \ \forall t \in T \ |f'(t,\theta)|\leq \lambda \text{ vs. }
H_1: \ \exists t \in T \  |f'(t,\theta)|> \lambda. 
\end{align}
Rejecting $H_0$ implies the existence of at least one time point at which a change in gene expression is detected that exceeds the critical value $\lambda$. It is important to note that $\lambda$ should be selected beforehand. Should any change in gene expression be of interest, $\lambda$ can be set to $0$. However, if only a larger change may be biologically relevant, $\lambda$ should be adjusted accordingly in consultation with experts. 
For instance, regarding the absolute difference of the model fit to $\log_2$-transformed counts at two time points $t_a$ and $t_b$, i.e. $|f(t_a,\theta)-f(t_b,\theta)|$, 
the corresponding threshold is often set to $\log_2(1.5)$ (see, e.g., \cite{WDM21}, \cite{KG21} and  \cite{MS22}), investigating whether the foldchange between $t_a$ and $t_b$ exceeds $1.5$. Based on this, a possible choice would be given by $\lambda=\frac{\log_2(1.5)}{\overline{t}}$ for the test regarding $f'$, representing the slope of a straight line that increases by $\log_1(1.5)$ between $t_1$ to $t_m$. 
Of note, $\overline{t}$ is interchangeable with any other time frame, e.g.  $\frac{1}{2}\overline{t}$. Then the question is whether there is an absolute change in $\log_2(1.5)$ over half of the study period.

Next, let $L^\alpha(t,\hat{\theta})$  denote a $(1-\alpha)$ lower simultaneous confidence band of $|f'(t,\theta)|$, that is
$$\mathbb{P}\{\forall t\in T: L^\alpha(t,\hat{\theta}) \leq |f'(t,\theta)|\}\geq 1- \alpha.$$
According to \cite{MS22}, rejecting $H_0$ in \eqref{hypotheses} if $L^\alpha(t,\hat{\theta})>\lambda$  for any $t\in T$, leads to an $\alpha$-level test, as
\begin{align*}
\mathbb{P}_{H_{0}}\{\exists t\in T: L^\alpha(t,\hat{\theta}) >\lambda \}
&\leq\mathbb{P}_{H_{0}}\{\exists t\in T: L^\alpha(t,\hat{\theta}) > |f'(t,\theta)| \}\\
&=1-\mathbb{P}_{H_{0}}\{\forall t\in T: L^\alpha(t,\hat{\theta}) \leq |f'(t,\theta)| \}\\
&\leq1-(1-\alpha)=\alpha. 
\end{align*}

The time period of significant change in gene expression corresponds to the set of all $t\in T$ for which $H_0$ in \eqref{hypotheses} can be rejected, i.e. $L^\alpha(t,\hat{\theta})>\lambda$, which we denote by
$\tilde{S}\coloneqq\{t|L^\alpha(t,\hat{\theta})>\lambda\}.$
\begin{figure}[ht!]
    \centering
\includegraphics[width=0.85\textwidth]{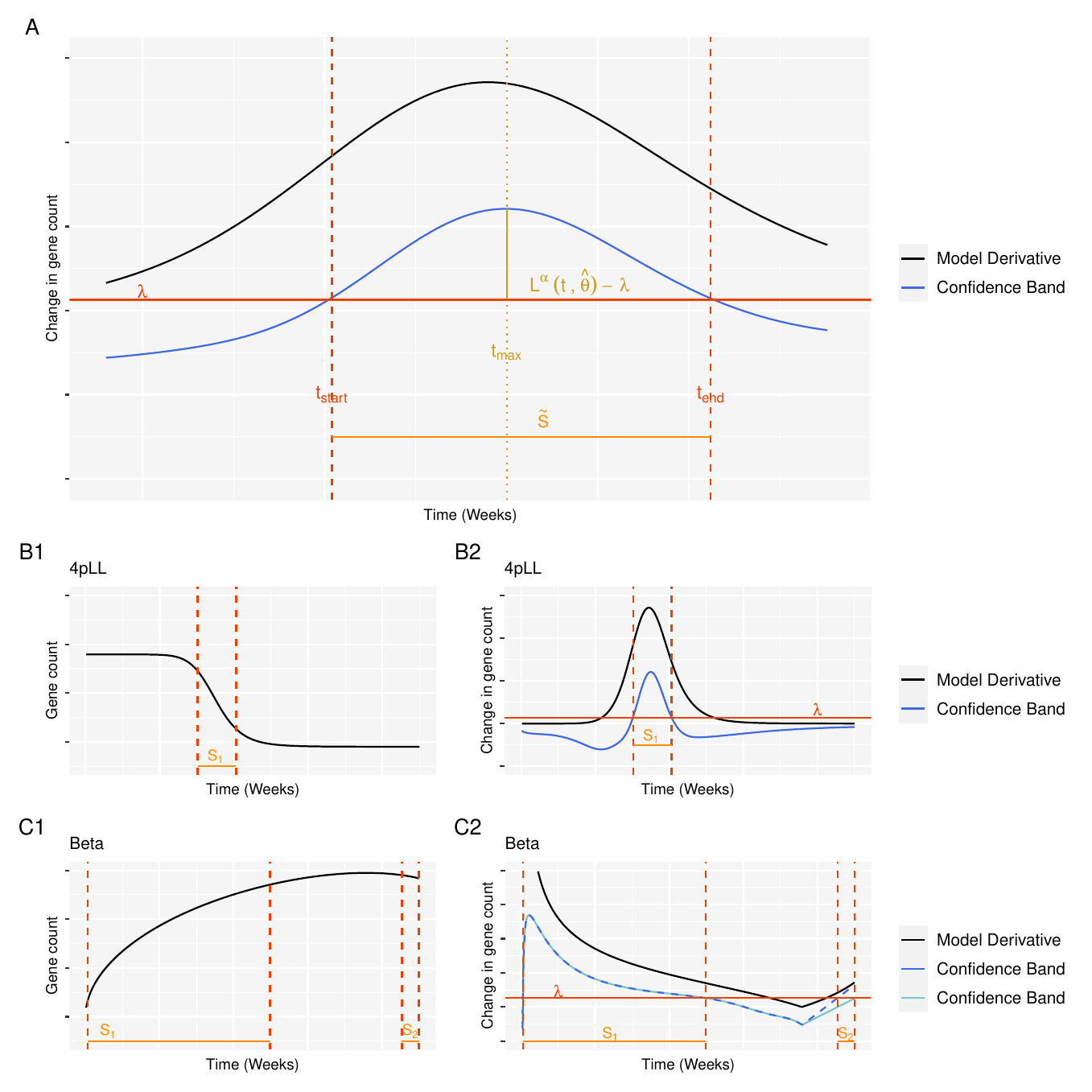}
    \caption{(\textbf{A}) Visualisation of the time period of significant change in gene expression $\tilde{S}$ (orange) with two time points of interest. The red dotted lines indicate the beginning and end of $\tilde{S}$. The yellow line indicates the time point of maximum change in expression. 
    $\tilde{S}$ is partitioned into coherent subsets if the gene expression curve follows a 4pLL model (\textbf{B}) or a beta model (\textbf{C}). \textbf{B1} and \textbf{C1} depict the fitted model, \textbf{B2} and \textbf{C2} the corresponding first derivative and  confidence bands such that $H_0$ in \eqref{hypotheses} is rejected. 
    }
    \label{exfigs}
\end{figure}
Figure \ref{exfigs} A depicts an illustrative example of a first derivative of a gene expression curve $f'$ (black) and its lower confidence band (blue). The orange line represents the time period $\tilde{S}$, during which the confidence band exceeds the threshold $\lambda$, indicated by the red line. 

While we can conclude significant changes in gene expression for all $t\in T$ which fulfill $L^\alpha(t,\hat{\theta})>\lambda$, some \textit{specific time points} might be of higher interest. An example are the beginning and end of a significant change. Precisely, these are the first time point $t_{start}\in T$ with $L^\alpha(t_{start},\hat{\theta})>\lambda$ 
and the last time point $t_{end}\in T$ with $L^\alpha(t_{end},\hat{\theta})>\lambda$. 
Figure \ref{exfigs} A indicates both $t_{start}$ and $t_{end}$ by the red vertical lines and thus also the period of significant change $\tilde{S}$. 
Another example is the time point of maximal change, indicated by the yellow vertical line in Figure \ref{exfigs} A, representing the time point $t_{max}\in T$ at which $L^\alpha(t,\hat{\theta})-\lambda$, $t\in \tilde{S}$ assumes its (locally) maximal value. 

These time points of interest are not necessarily unique. In case of a non-monotonous function $f$ (e.g. beta model), the confidence band can dip below the threshold $\lambda$ and than exceed it again after a time period of no significant change. Therefore, we can partition $S$ into ascending coherent subsets
$\tilde{S}=\big(S_{1},\dots,S_{s}\big),$ meaning that for all $j=1,\dots,s-1$,  it exists a $t'\in T$ between $S_j$ and $S_{j+1}$ such that $L^\alpha(t',\hat{\theta})<\lambda$ and $\max(S_j)<\min(S_{j+1})$. 
We refer to Figure \ref{exfigs} B and C for a visual representation. Figure \ref{exfigs} B depicts a gene expression curve that follows a 4pLL model on the left and its first derivative with a possible confidence band on the right. Here, $H_0$ is rejected and the time period of significant change in gene expression is coherent. Then $\tilde{S}$ equals $S_1$.  Figure \ref{exfigs} C depicts a beta model in the left and its first derivative with two possible confidence bands on the right. The solid light-blue line represents the scenario in which the decline in gene count in the latter part of the study period is not statistically significant.  The dotted blue line on the other hand represents the scenario in which the aforementioned decline is significant. Here, $\tilde{S}$ splits into the coherent sub-periods $S_1$ and $S_2$. 
Each $S_j$, $j=1,\dots,s$, has a unique set of the aforementioned time points of interest. In the following, we will focus on $t_{start}$ and $t_{end}$ as an example. Therefore, the set of times to search for is given by
\begin{align}
    \{t^{(j)}_{start}=\min(S_j)\mid j=1,\dots,s\}\cup\{t^{(j)}_{end}=\max(S_j)\mid j=1,\dots,s\}. \label{S}
\end{align}
The precise estimation of these quantities depends on the accuracy of the estimation of $L^\alpha(t,\hat{\theta})$, which is described in the next section.



\subsection{Estimating a confidence band of $\mathbf{|f'(t)|}$}\label{subsec:cfb}
 In order to  estimate a lower simultaneous confidence band $L^\alpha(t,\hat{\theta})$  of $|f'(t,\theta)|$ the method developed in \cite{MS22} is adjusted to the first derivative. Therefore, we define
$$L^\alpha(t,\hat{\theta}) :=|f'(t,\hat{\theta})|-c\hat{\sigma}_{|f'(t,\hat{\theta})|}
$$
and estimate the critical value $c$, such that 
$$
\mathbb{P}\bigg\{\max_{t\in T}\frac{|f'(t,\hat{\theta})|-|f'(t,\theta)|}{\hat{\sigma}_{|f'(t,\hat{\theta})|}}\leq c\bigg\}=1-\alpha.$$


To estimate both $\hat{\sigma}_{|f'(t,\hat{\theta})|}$ and $c$, a two-step parametric bootstrap procedure inspired by the classic bootstrap-t method (see \cite{ET94})  has been proposed in Algorithm 1 by \cite{MS22} and is applied here by substituting the difference for the first derivative.
In short, $\hat{\theta}$ is estimated from the original data set and used to generate $B_1$ bootstrap samples. This is done by randomly drawing iid. errors $\epsilon_{p,q}^*\sim N(0,1)$ and applying the formula $$y_{p,q}^*=f(t_p,\hat{\theta})+\hat{\sigma}\epsilon_{p,q}^* \ 
 \text{ for } \ p=1,\dots,m, 
 \ q=1,\dots,n_p.$$
 These can be considered the first level of the bootstrap. 
 For each of the generated samples,  $\hat{\theta}^*_l$, $l=1,\dots,B_1$,  are estimated and used to generate another $B_2$ bootstrap samples, which constitute the second level of the bootstrap. Then, $\hat{\sigma}_{|f'(t,\hat{\theta})|}$ is estimated from the first bootstrap level by calculating the empirical variance of the sample $|f'(t,\hat{\theta}^*_1)|,\dots,|f'(t,\hat{\theta}^*_{B_1})|$. Similarly, $\hat{\sigma}_{|f'(t,\hat{\theta}^*_l)|}$, $l=1,\dots,B_1$, is calculated for each of the generated first-level bootstrap samples using the second-level bootstrap samples. With these values $$D^{*,l}:=\max_{t\in T}\frac{|f'(t,\hat{\theta}_l^*)|-|f'(t,\hat{\theta})|}{\hat{\sigma}_{|f'(t,\hat{\theta}_l^*)|}}$$ can be calculated  for each $l=1,\dots,B_1$. The respective empirical $(1-\alpha)$-quantile of the distribution yields an estimate of the critical value $c$. A visual representation of this process is provided in the blue rectangle in Figure A (see supplementary material).

\subsection{Confidence intervals for the time points of interest}\label{subsec:cfi}

In the previous sections we obtained time points of significant change in gene expression. However, as estimators, they are subject to uncertainty. Thus, the construction of confidence intervals for these time points is of huge practical importance. 
The method proposed is an extension to the two-step bootstrap described in Section \ref{subsec:cfb}. 
Precisely, another bootstrap level is added in advance, and the two-step procedure described above is then applied to each sample. This provides an estimated lower confidence band for each of the generated data sets, allowing calculation of the time frames of significant change in gene expression and the time points of interest. In the following analysis, only those cases with the same number of coherent subsets as the original data set are considered. 
The corresponding time points of interest from all bootstrap samples are combined (e.g. all $t^*_{start}$ and all $t^*_{end}$, where the '*' denotes the bootstrap data) and used to estimate the empirical $\frac{\alpha}{2}$- and $(1-\frac{\alpha}{2})$-quantiles , say $q_\frac{\alpha}{2}$ and $q_{(1-\frac{\alpha}{2})}$ providing percentile confidence intervals for the time points of interest of the original data set that can be expressed as $[q_\frac{\alpha}{2},q_{(1-\frac{\alpha}{2})}]$. Details can be found in the Supplemental material and the Appendix: Figure A (see supplementary material) depicts a visualization of this concept, while Algorithm \ref{algcit} summarizes the procedure.

\section{Simulation study}\label{sec:sim}
\subsection{Setting and data generation}\label{sec:sim:datagen}
\begin{figure}[ht!]
        \centering
        \includegraphics[width=0.85\textwidth]{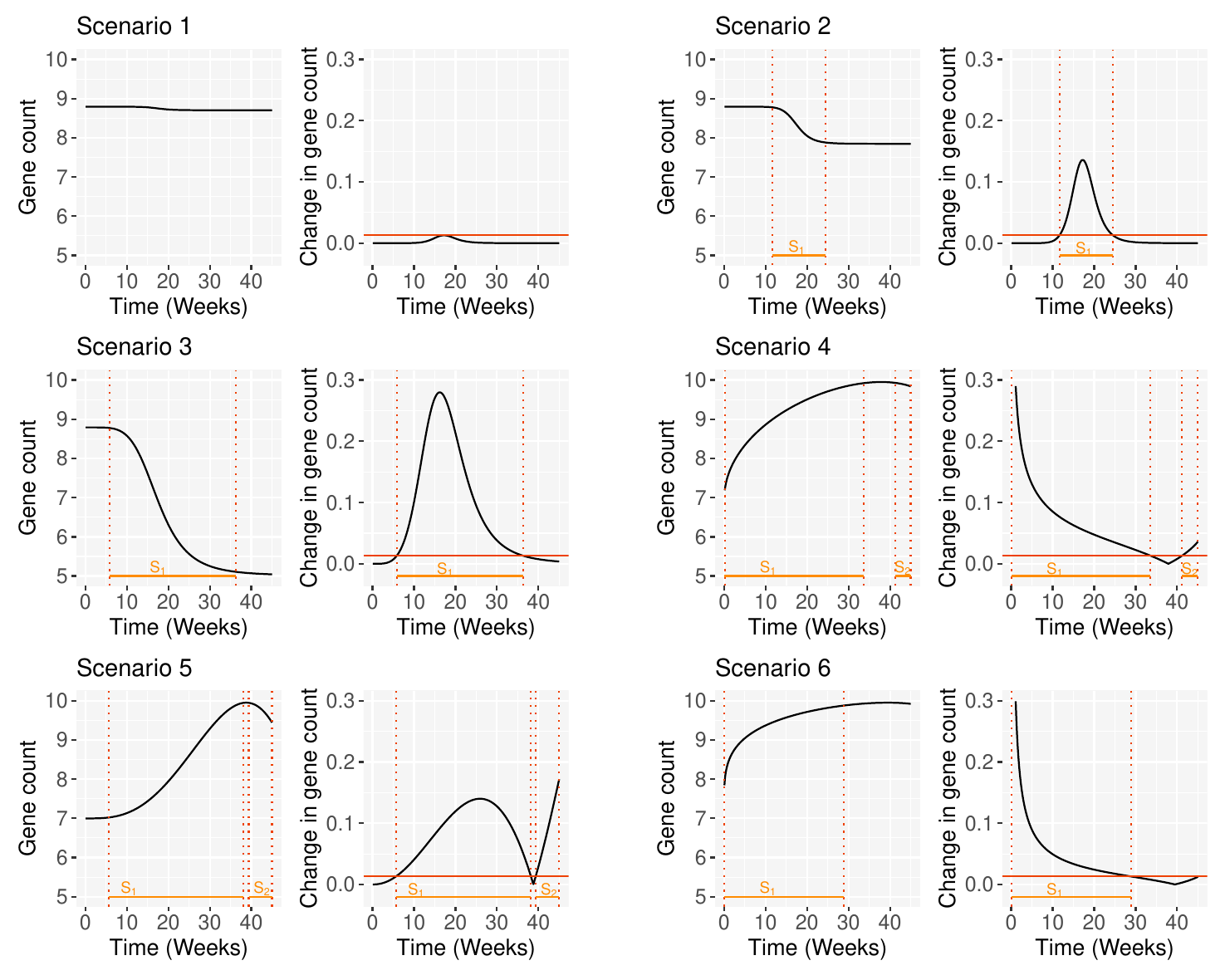}
        \caption{Visualization of the different scenarios. The horizontal red line indicates the relevance threshold $\lambda=\frac{\log_2(1.5)}{45}$. Each case depicts the true model on the left and the true first derivative on the right. The vertical red lines mark the time frame of relevant change in regards to $\lambda$. 
        }
        \label{Visscen}
    \end{figure}

In order to evaluate the performance of our newly proposed method, a simulation study was conducted. Motivated by \cite{MS22} and \cite{KG21}  we focused on the 4pLL and the beta model. The simulation settings were inspired by the Western diet mice trial discussed in Section \ref{sec:casestudy}. Precisely, we used the parameters obtained from the expression of the genes 'Cd163' (4pLL model, see Section \ref{sec:casestudy:individual} for details) and 'Fam83a' (beta model) 
in WD-fed mice. This resulted in the six basic scenarios, which are summarized in Table \ref{tab:scen} and 
depicted in Figure \ref{Visscen}. 

Scenarios 1-3 were generated using the 4pLL model, while the latter three were generated with the beta model.
In Scenario 1, there is no relevant change in gene expression respective to the threshold $\lambda=\frac{\log_2(1.5)}{45}\approx 0.0130$. Therefore, $H_0$ in \eqref{hypotheses} cannot be rejected. 
In all other scenarios, there is a change in expression larger than $\lambda$, and, therefore, these scenarios correspond to the situation under the alternative.  
Scenario 4 reflects the unaltered parameters as estimated. Scenarios 2, 3, 5 and 6 were based on the real examples, but changed to reflect a different slope, see Table \ref{tab:scen}.
\begin{table}[t]
    \centering
        \caption{\raggedright Overview of simulation scenarios.}
    \begin{tabular}{l||l|l|l|l}
    Sce.& Short name& Model& Parameters & Description\\
    \hline
       1&No relevant change&4pLL& $a=8.791$, $b=-0.089$ &  Does not exceed $\lambda=0.0130$ \\
       &&&$c=17.589$, $h=10.000$&Reflects the margin of $H_0$.\\
       \hline
       2&Small Jump&4pLL&$a= 8.791$, $b=-0.946$&Exceeds $\lambda=0.0130$\\
       &&&$c=17.589$, $h=10.000$&$\tilde S=(11.7,24.5)$\\
         \hline
       3&Large Jump&4pLL&$a=8.791$, $b=-3.783$& Exceeds $\lambda=0.0130$\\
       &&&$c=17.589$, $h=5.000$&$\tilde S=(5.9,36.3)$\\
         \hline
       4&Dip&beta&$a=6.997$, $b=2.952$& Exceeds $\lambda=0.0130$ twice\\
       &&&$\delta_1=0.506$, $\delta_2=0.215$&$S_1=(0,33.6)$, $S_2=(41.2,45)$\\
         \hline
       5&Dip alternative&beta&$a=6.997$, $b=2.952$& Exceeds $\lambda=0.0130$ twice\\
       &&&$\delta_1=3.286$, $\delta_2=1.290$&$S_1=(5.7,38.1)$, $S_2=(39.4,45)$\\
         \hline
       6&No Dip&beta&$a=6.997$, $b=2.952$&Exceeds $\lambda=0.0130$\\
       &&&$\delta_1=0.228$, $\delta_2=0.084$&$\tilde S=(0,28.9)$
    \end{tabular}
    \label{tab:scen}
\end{table}


The observations were simulated at seven time points ($0$, $3$, $9$, $15$, $21$, $27$, $33$, $39$ and $45$ weeks) in accordance with the feeding times observed in the Western diet mice trial, following the beginning of the study. A total of $47$ data points were simulated consisting of $5$ mice at the first $5$ time points and $4$ and $8$ mice at $39$ and $45$ weeks, respectively.

In order to gain a comprehensive understanding of the impact of the standard deviation, five distinct levels for each of the six basic scenarios were considered: 'small','mid-small', 'medium', 'mid-large' and 'large'.  
The 'medium' standard deviation of the realistic Scenario 4 is as estimated from the real world data. For all other scenarios, the 'medium' standard deviation was calculated via linear transformation with regard to the respective maximum absolute differences in gene expression 
compared to the realistic example of the same model type. 
Regarding the parameterization introduced in Section \ref{subsec:model}, the maximum absolute value corresponds to the parameter $b\in\theta$. Accordingly, we designate $b_{sce}\in\theta_{sce}$  as the maximum absolute difference of the scenario and $b_{real}\in\theta_{real}$ as the maximum absolute difference estimated from the example.
With this, we obtained the transformed standard deviation by
$\sigma_{sce,medium}=\frac{b_{sce}}{b_{real}}\sigma_{real}$,
where  $\sigma_{real}$ the estimated value from the real world data. 
From these 'medium' values of standard deviation, the 'small', 'mid-small', 'mid-large' and 'large' levels were calculated by multiplying with $0.5$, $0.75$, $1.5$ and $2$, respectively, see Table \ref{tab:sds} for details.
\begin{table}[ht]
           \centering
        \caption{\raggedright Levels of standard deviation used for the simulation study. The standard deviations marked with an asterisk ($^*$) are approximately half the size of the maximum change in gene expression.  }
    \begin{tabular}{l||l|l|l|l|l}
        Scenario & $\sigma$ small &$\sigma$ mid-small &$\sigma$ medium&$\sigma$ mid-large  &$\sigma$ large  \\
        \hline
         1&$0.014$&$0.021$&$0.028$&$0.042$&$0.056^*$\\
         2&$0.149$&$0.223$&$0.297$&$0.446$&$0.595^*$\\
         3&$0.595$&$0.892$&$1.190$&$1.784^*$&$2.380^*$\\
         4&$0.283$&$0.424$&$0.566$&$0.849$&$1.132$\\
   5&$0.283$&$0.424$&$0.566$&$0.849$&$1.132$\\
   6&$0.283$&$0.424$&$0.566$&$0.849$&$1.132$\\
    \end{tabular}
    \label{tab:sds}
\end{table}

Given either the model function, $f$, of the beta model or the 4pLL model and the values of $\theta$ and $\sigma$, the gene expression data was simulated by drawing iid. errors $\epsilon^{sim}_{p,q}\sim N(0,1)$ and applying the formula $$y_{p,q}^{sim}=f(t_p,\hat{\theta})+\hat{\sigma}\epsilon^{sim}_{p,q} \text{ for } p=1,\dots,m, \ q=1,\dots,n_p.$$ 
For the two-step bootstrap procedure we chose $B_1=500$ and $B_2=25$ repetitions. Each procedure was repeated in $1,000$ simulation runs. Outcomes of the simulation study were the number of rejections of $H_0$ in \eqref{hypotheses} and the identified time frame of significant change in gene expression.
The majority of the code was run on \texttt{R} version 4.3.2 using a Windows 11 x64 PC. The simulation scenarios were run on a HPC-cluster employing \texttt{R} version 4.3.1. The runtime of each simulated study on a commercially available Windows PC is approximately 5 minutes, with a variability of a few minutes depending on the number of CPU cores.
    
    \subsection{Results}\label{sec:ressim}
    \subsubsection{Rejection of $\mathbf{H_0}$}
\begin{figure}
    \centering
    \includegraphics[width=.9\textwidth]{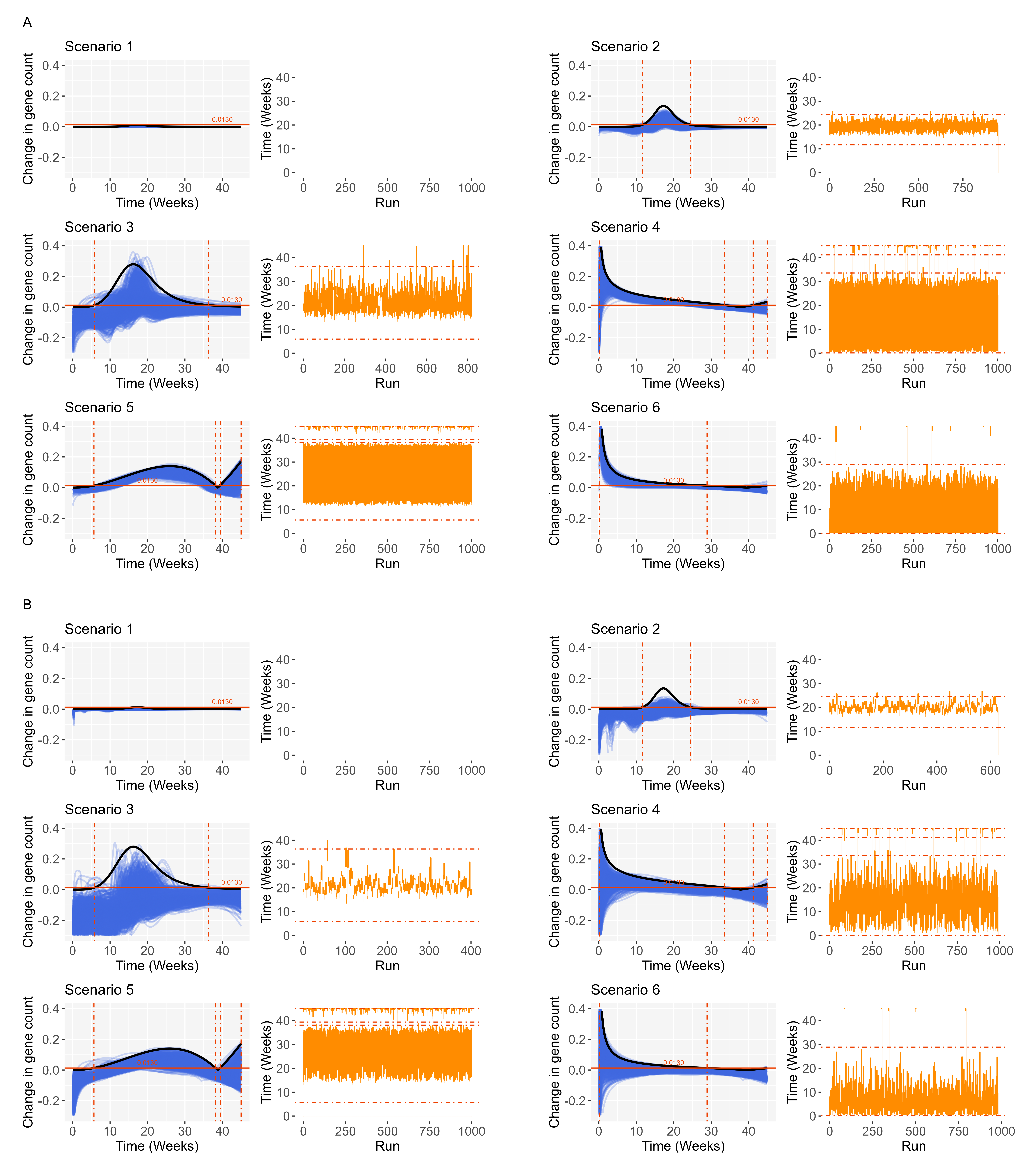}
    \caption{Results of the six simulation scenarios for a mid-small (A) and mid-large (B) standard deviation. The left column depicts the true first derivative and the estimated confidence bands. The right column depicts the estimated time frames of significant change in gene expression for each run.}
    \label{resmsmall}
\end{figure}

    Table \ref{h0rej}  shows the number of rejections of $H_0$ in $1,000$ runs for all configurations. The left column in Figure \ref{resmsmall} depicts the estimated confidence bands for a mid-small and mid-large standard deviation. We refer to the supplementary material Figures D-F for the other cases. 

\begin{table}[htbp]
  \centering
  \caption{\raggedright Number of rejections of $H_0$ in $1,000$ runs.}
    \begin{tabular}{l||l|l|l|l|l|l}
  $\sigma$ level  &Sce. 1 & Sce. 2 & Sce. 3 & Sce. 4 & Sce. 5 & Sce. 6 \\
        \hline
   small $\sigma$& 0     & 999   & 1000  & 1000  & 1000  & 1000 \\
   mid-small $\sigma$& 0     & 946   & 818   & 1000  & 1000  & 1000 \\
medium $\sigma$&    0     & 830   & 514   & 1000  & 1000  & 1000 \\
   mid-large $\sigma$& 0     & 627   & 401   & 989   & 999   & 978 \\
   large $\sigma$& 0     & 372   & 303   & 814   & 957   & 780 \\
    \end{tabular}%
  \label{h0rej}%
\end{table}%

In Scenario 1,  no false rejection of $H_0$ was observed in any of the cases, indicating a type I error of 0. This scenario corresponds to the margin of the null hypothesis in \eqref{hypotheses}, here given by $\lambda=0.0130$. The conservatism of the test decision even for a large standard deviation may therefore be partially attributed to the effect being very small and an overall tendency to slightly underestimate the time frame of significant change in expression discussed in the next subsection.

In general, our proposed method shows high power when the beta model is the true model. Up to a medium standard deviation, $H_0$ is always rejected for Scenarios 4, 5 and 6. Even in the case of a large standard deviation, the power remains at about $80\%$ for Scenarios 4 and 6 and above $95\%$ for Scenario 5. 
Accordingly, the estimated confidence bands 
show a low variability.

For the remaining scenarios generated with the 4pLL model -- Scenario 2 and 3 -- we observe a large influence of the standard deviation. In particular, significant changes in gene expression were observed for a mid-large or large variability below $650$ and up to $303$. This influence seems to be more pronounced for the larger jump in gene expression simulated in Scenario 3. In this case, only $514$ of significant changes were detected for a medium standard deviation. 

It is noteworthy that the (mid-)large standard deviations (see Table \ref{tab:sds}) employed in the generation of Scenarios 2 and 3 is considerably large compared to the absolute change in gene expression. Naturally, this leads to a high variability in the generated simulation studies (see Figure B in the supplementary material). However, we observe that in the case of the 4pLL model, the translation of a high degree of variability on the count scale leads to an extreme degree of  variability on the scale of the change in gene expression considered when applying the first derivative (see C in the supplementary material). It is important to interpret the results presented in light of these considerations. It is also important to note that the same effect occurs when generating the first and second level bootstrap runs. This is likely to have an impact on the estimation process, especially if the data set has a high degree of variability.

For Scenarios 2 and 3 we note a visible crater in the estimated confidence bands around the week where the true first derivative reaches its maximum, which corresponds to the inflection point of the model function. 
While the extent is likely to be at least partially attributable to the size of the standard deviation, the trend is discernible across all cases. Given that the effect appears to be more pronounced in Scenario 3 than in Scenario 2, it is plausible that its magnitude may correlate with the steepness of the true model function in the area around its inflection point. This conclusion is supported by the absence of a comparable crater for Scenario 5, the only scenario with an inflection point generated using the beta model, which is markedly less steep than Scenarios 2 and 3. 

\subsubsection{Estimated time frame of significant change in gene expression}
While the number of rejections of $H_0$ is largely satisfactory, a reliable detection of the time frame of significant change in gene expression is of greater interest. 
Therefore, we considered the bias and variance of each start/end time point of the estimated time frame of significant change in gene expression.
Furthermore, we examined the number of runs in which the true number of coherent subsets of $\tilde S$ was correctly identified. We refer to Table A in the supplementary material for a summary of all results.
As anticipated, the true number of coherent subsets was consistently detected for Scenarios 2 and 3 if $H_0$ was rejected. Given the monotonicity of the 4pLL model, the time frame of significant change in expression is, by definition, continuous. However, as described above, a crater was observed in some estimated lower simultaneous confidence bands. In a small number of cases, this resulted in the identification of two disjointed subsets. Overall, the number of these cases is less than $12$ for Scenario 2. In case of Scenario 3, the same is observed except for a mid-small and medium standard deviation, which respectively fall within the ranges of $200$ and $70$ cases.

While $H_0$ was reliably rejected for Scenarios 4 and 5, the second time frame of significant change in expression was not always detected. The smaller dip in Scenario 4 was only identified  approximately $120$ times when the standard deviation was small and approximately $40$ times otherwise. The more pronounced dip in Scenario 5 was detected $999$ times for a small standard deviation, $671$ times for a medium standard deviation and $260$ times for a large standard deviation. However, it should be noted that in both scenarios, the maximum value in gene count is reached around week $39$, which is the second to last sample time point. Therefore, the detection of the dip is contingent only on the last measurement. 
In Scenario 6, a dip at the end was incorrectly identified less than $15$ times in all cases.

The right columns in Figures \ref{resmsmall} depict the estimated time frames of significant change  in gene expression for each of the $1,000$ simulation runs for the mid-small and mid-large standard deviations, respectively. We refer to Figures and D-F in the supplementary material for the other cases. The true values are indicated by the red lines. In general, we note that the estimated start of the time frame of significant change in expression displays a positive bias for all scenarios, whereas the estimated end displays a negative bias. Therefore, the time frame is overall slightly underestimated. The bias depends largely on the standard deviation and the slope of the first derivative. Scenario 6 exhibits the smallest bias for the onset of the significant change at the beginning of the study, never exceeding $0.375$ weeks. Scenario 4 is similarly flat. Its bias never exceeds $2.379$.
Scenario 5 is characterised by a steeper slope and exhibits a noticeably larger bias. The range is from $3.956$ weeks for a small standard deviation to $8.865$ weeks for a large standard deviation. For Scenarios 2 and 3 -- disregarding the crate when detected -- we observe similar results.
The influence of the slope seems to have a reverse effect on the bias of the end of the identified time frame of significant change. In particular, flat scenarios, such as Scenario 6, demonstrate a noticeably larger bias for all levels of standard deviation. In addition, the bias is overall larger than for the start of the significant change, with a range of $-1.225$ to $-23.591$ weeks. 

In Scenarios 4 and 5,  a second time frame of significant change in gene expression was simulated. Since in both cases it is cut off at the end of the study, we consider only the starts of the second coherent subsets of $\tilde S$, respectively. Again, we note a positive bias, which ranges from almost $2$ weeks to nearly $3.5$ weeks.
Finally, the variance shows a similar behavior to the bias.

 	\section{Case study} 
  \label{sec:casestudy}
 In this section, three possible applications of our method are presented. For illustration, we use data from the Western diet mice study (see \cite{WDM21}). A total of $79$ male mice were fed a Western (WD) or standard (SD) diet for a maximum of $48$ weeks to assess the effect of a high fat diet on the development of non-alcoholic fatty liver disease. Up to $8$ mice per diet were sacrificed at $3, \  6, \ 30, \ 36, \ 42$ and $48$ weeks, and WD-fed mice were additionally sacrificed at $12, \ 18$ and $24$ weeks. RNA was extracted from frozen liver tissue and RNA-seq analysis was performed. 
Here, we use the pre-processed data set, that was generated using Salmon (see \cite{PD17}) and the \texttt{R} packages tximeta (see 
 \cite{LS20}) and DESeq2 (see \cite{LH14}). The count data was normalized and $\log_2$-transformed using the \textit{vst()} function (see \cite{Tib88}, \cite{HH03} and \cite{AH10}) from the DESeq2 package. As the initial samples were collected at $3$ weeks, the beginning of the study at $t=0$ in the subsequent analysis represents these initial data points.

The first example is an application to a single gene. In the second example, our method is used to assess the effect of SD vs. WD on gene expression, using a gene that shows differential expression in both cases as an example. In the third example, we select a pool of nearly $10000$ genes and use our method to identify those that change significantly between 10 and 25 weeks. A Gene Ontology (GO) enrichment analysis is then performed on the identified genes.
In all cases, the two-step bootstrap procedure was implemented with $B_1=500$ and $B_2=25$ bootstrap runs, while the three-step bootstrap procedure was executed with an additional $B_3=500$.

\subsection{Application to an individual gene}\label{sec:casestudy:individual} 
'Cd163' is a gene found in both mice (see \cite{Cd163mice}) and humans (see \cite{Cd163human}). The soluble form of the encoded molecule was found to be a predictor for both fibrosis and hepatocellular carcinoma development in nonalcoholic steatohepatitis in humans, as reported in \cite{KN23}. In particular, decreased levels were observed in patients with NAFLD who exhibited alleviated fibrosis and inflammation. In \cite{WDM21},
the gene has been identified as a RJG. This implies that the gene in question is only deregulated after a specific period of time, in this case, after $24$ weeks.
In a preceding model selection step, both the 4pLL and the beta model were fitted to the gene expression data of the WD-fed mice. The AIC for the 4pLL model was $90.4$, while that for the beta model was $98.2$. Therefore, the 4pLL model was used for further analyses.

\begin{figure}
    \centering
    \includegraphics[width=0.9\textwidth]{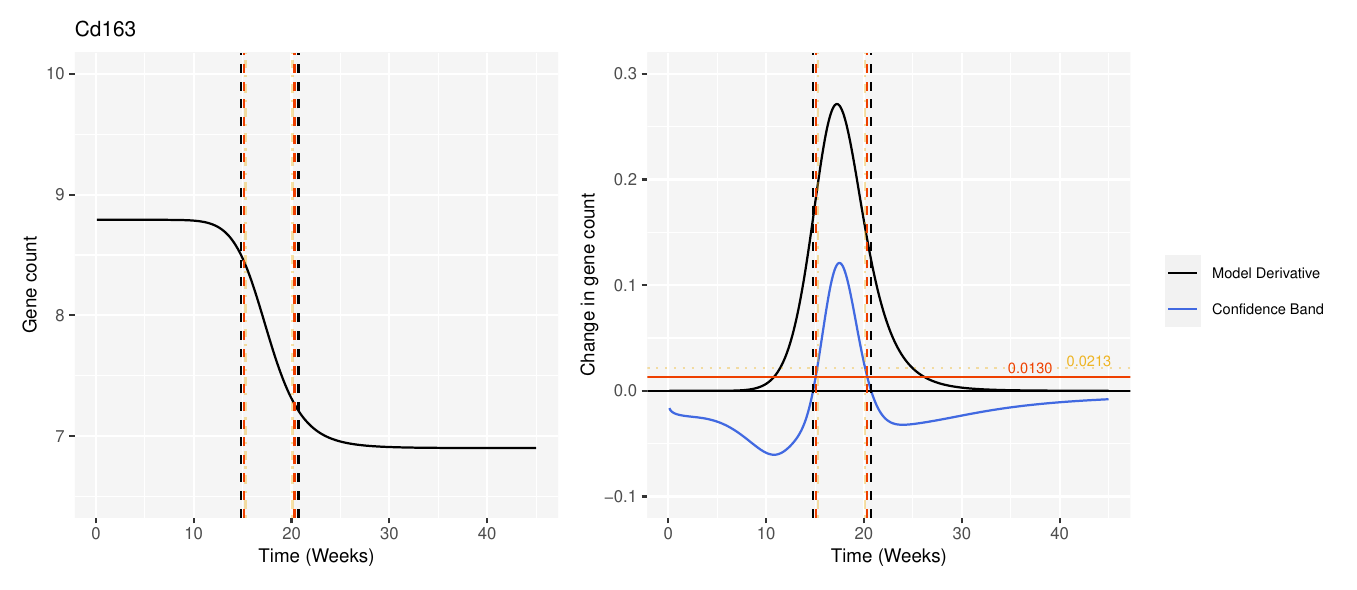}
    \caption{On the left, the fit of the 4pLL model to the normalized gene count of the gene ''Cd163'' over time is depicted. On the right, the first derivative and the estimated confidence band of the model fit are displayed. The horizontal red line indicates $\lambda=\frac{\log_2(1.5)}{45}$. The vertical dashed red lines indicate the corresponding time frame of significant change in gene expression. The yellow colored lines correspond to an alternative choice of $\lambda=\frac{\log_2(1.5)}{27.5}$. The black lines represent the case $\lambda=0$.}
    \label{cc163}
\end{figure}

The left panel of Figure \ref{cc163} depicts the estimated gene expression curve. In accordance with the identification as a RJG, it appears that approximately $10-12$ weeks pass, before the gene becomes down-regulated. Following an additional $10-12$ weeks, the deregulated expression appears to reach a constant value. Overall the gene changes by a value of $ -1.891$. 
In the right panel of Figure \ref{cc163}, the corresponding first derivative and the estimated confidence band are depicted. The various horizontal lines represent different potential choices for the threshold $\lambda$, while the corresponding vertical lines represent the respective time frames of significant change in gene expression.
The black line represents the case where $\lambda=0$, while the red line corresponds to the aforementioned case  $\lambda=\frac{\log_2(1.5)}{45}=0.0130$ (see Section \ref{subsec:test}). Similarly, a third threshold was calculated, indicated by the yellow line, which corresponds to a change of $\log_2(1.5)$ in gene expression over the span of half the study. This third threshold is given by $\lambda=\frac{\log_2(1.5)}{27.5}=0.0213$.


We observe that $H_0$ is rejected for all thresholds. 
In the following, we will present the analysis for $\lambda=\frac{\log_2(1.5)}{45}$ as an illustrative example.
The period during which the estimated significant changes in gene expression occur begins  at $15.1$ weeks ($95\%$ CI $\left[13.2, 17.8\right]$) and ends at $20.3$ weeks ($95\%$ CI $\left[17.4, 23.5\right]$). Consequently, we can conclude the classical RJG form of a delayed deregulation. The change is observed over a period of $5.2$ weeks, with a plateau following the initial 'jump' in activity. Over the course of the $5.2$ week period, the expression undergoes a change of $1.189$. Therefore, we can account for $62.883\%$ of the observed maximal change via a significant effect.

For comparison, the other cases indicate a time period of significant change in gene expression between weeks $14.8$ to $20.7$ ($\lambda=0$, black) and $15.3$ to $20.1$ ($\lambda=0.0213$, yellow), respectively.
Consequently, while the 'Cd163' gene was also identified as a RJG, the estimated time frame of deregulation is set earlier than reported in \cite{WDM21} for all thresholds.

\subsection{Aging}
For this application we consider a gene that deregulates in both diet groups and compare the estimated time frames of significant change in gene expression. The biological question inspiring this example is whether the WD accelerates or decelerates, and thereby 'ages' biological processes, which also occur in SD-fed mice. The chosen example, 'Dbp', is a protein encoding gene detected in humans \cite{Dbphuman} and mice \cite{Dbpmice} associated with the circadian rhythm.
A preliminary model selection step indicated a superior fit of the 4pLL model to the gene expression data for both diet groups. Consequently, it was utilised in the subsequent analysis.

\begin{figure}
    \centering
    \includegraphics[width=0.85\textwidth]{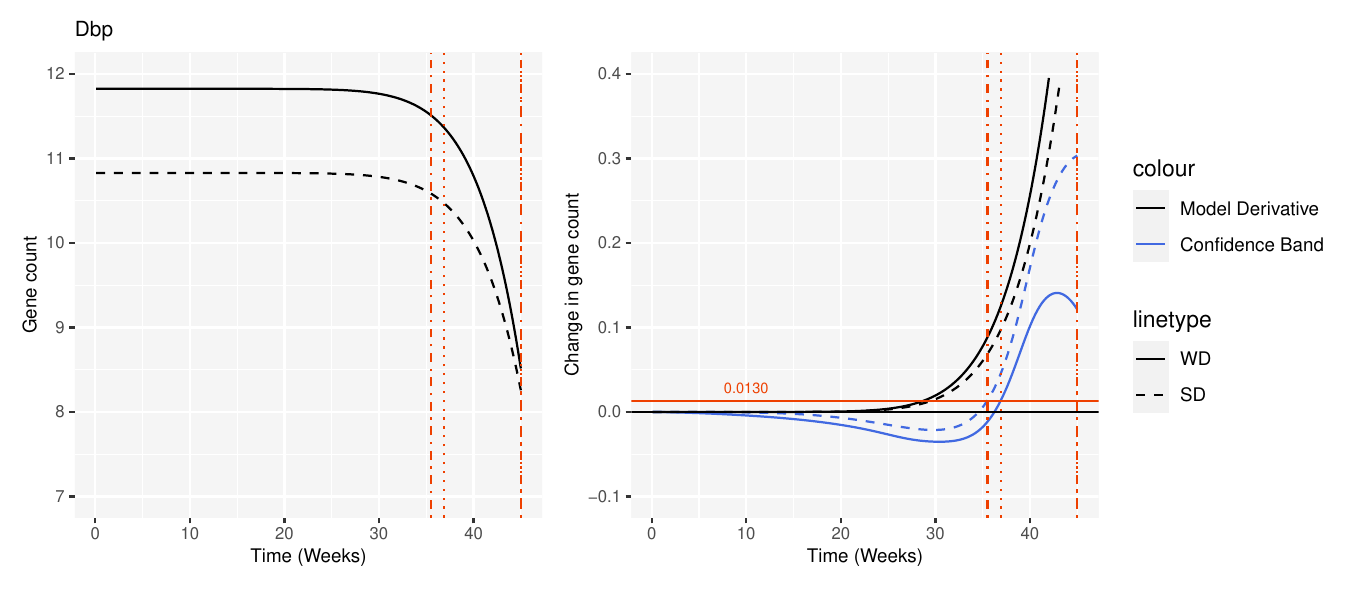}
    \caption{A comparison of the SD versus the WD, exemplary for the gene ''Dbp''. The gene counts, as modelled by the 4pLL model, are depicted on the left. On the right, the first derivatives and their estimated confidence bands are shown. The red dotted line outlines the time period of significant change in gene expression for the WD, while the dot-dashed line outlines the analogous period for the SD.}
    \label{Dbp_comparison}
\end{figure}

The left panel of Figure \ref{Dbp_comparison} depicts the modeled gene expression of both the SD and the WD groups. The gene count in SD-fed mice initially appears to be smaller and both genes seem to undergo similar deregulation at approximately the same time and converge to the same value. On the right, the corresponding first derivatives and the estimated confidence bands are displayed. The dotted line represents the time period of significant change in gene expression for the WD-fed group, while the dot-dashed line outlines the time period for the SD-fed mice.
Although the first derivative of the WD-fed group appears to increase earlier, the confidence band of the SD group exceeds the threshold $\lambda=\frac{\log_2(1.5)}{45}=0.0130$ earlier. 
The estimated time frame of significant change in gene expression in WD-fed mice ranges from $36.9$ weeks ($95\%$ CI $\left[33.4, 38.4\right]$)  to the end of study at $45$ weeks ($95\%$ CI $\left[40.5, 45.0\right]$) , indicated by the red dotted lines. Over the span of $8.1$ weeks the gene count undergoes an absolute change of $2.213$. 
Similarly, the estimated time period during which significant change in gene expression in SD-fed mice occurs begins at $35.5$ weeks ($95\%$ CI $\left[32.2, 36.3\right]$) and ends with the end of study at $45$ weeks ($95\%$ CI $\left[43.2, 45.0\right]$). The gene count undergoes a change of $2.996$ over the span of $9.5$ weeks.
Overall, both genes show very similar results. The estimated confidence band for the SD-fed mice exceeds the threshold $\lambda=0.0130$ about $-1.4$ weeks earlier and the  estimated $95\%$ confidence interval for this difference is given by $\left[-4.9,  2.0\right]$. 


\subsection{GO enrichment analysis}\label{casestudy:GO}
For this example we applied our proposed method to identify a subset of differentially expressed genes that deregulate during a pre-selected time period ($10$ to $25$ weeks). In a subsequent step, we conducted a GO enrichment analysis on these genes.  This means, we examined if a significant number of the identified genes were assigned to pre-specified GO groups (see \cite{Con01}), in which the genes were grouped according to their connection to  biological functions. This approach would permit a practitioner to draw inferences regarding the time frame within which specific biological processes occur.

To achieve this, we used the differential gene expression analysis as performed in \cite{WDM21} using the DESeq2 package to pre-select the subset of genes that are estimated to undergo significant changes in expression between week $0$ (time on study) in SD-fed mice (SD0) and week $45$ in WD-fed mice (WD45). Precisely, the pre-processed data set was created by applying a generalized linear model with one factor, a combination of the diet and the week. The SD0 group was used as reference. 
To account for the potentially high variability of the counts, we used this as basis and performed a shrinkage of the effect size WD45 vs. SD0 as described in \cite{ZI18} by applying the \textit{lfcShrink()} function from the DESeq2 \texttt{R} package. Afterwards we 
identified all genes with an adjusted p-value (see \cite{BH95}) smaller than $0.05$ when comparing the expression of SD-fed mice at $0$ weeks to the expression of WD-fed mice at $48$ weeks.

\begin{figure}
    \centering
    
    \includegraphics[width=0.7\textwidth]{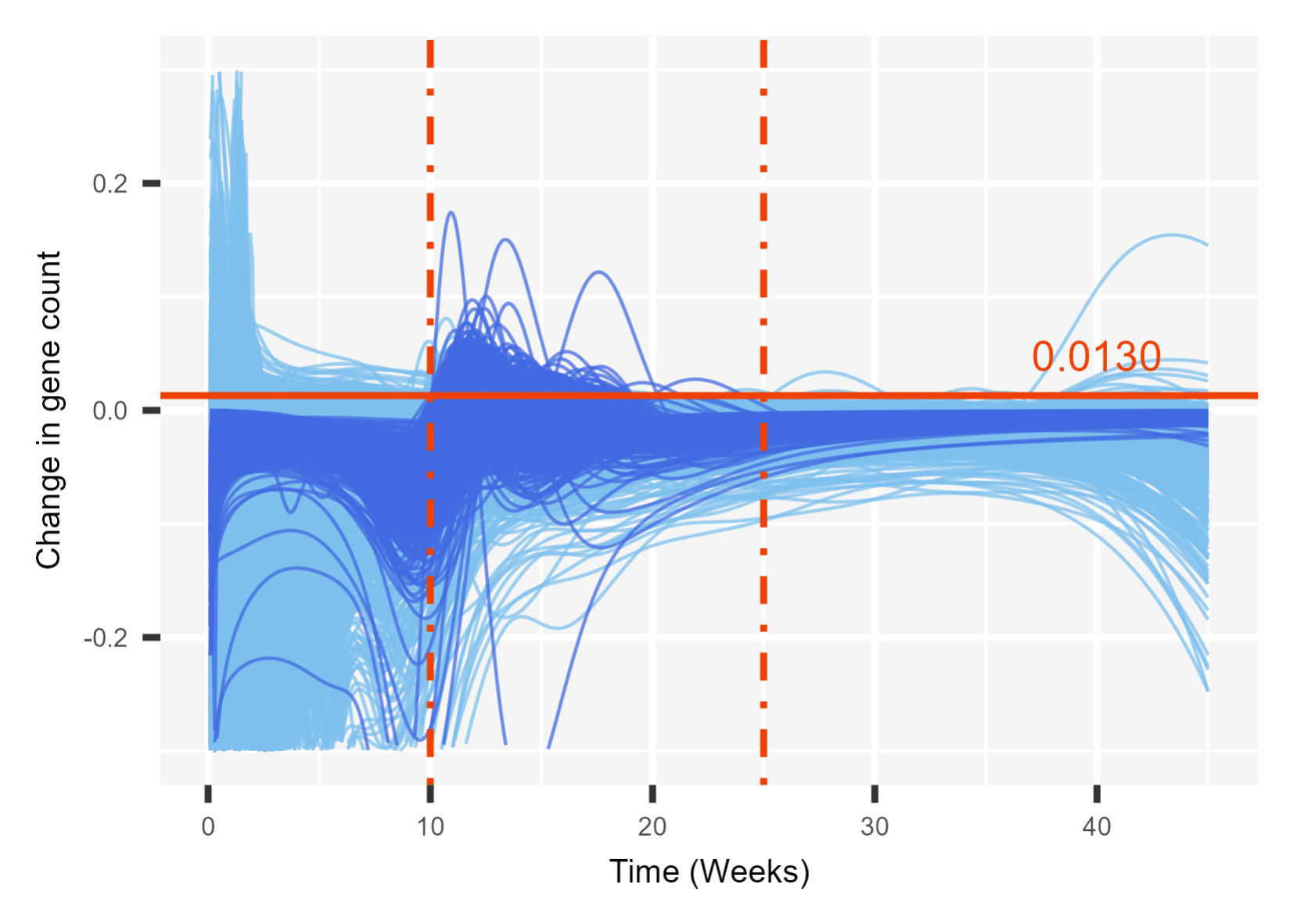}
    \caption{Estimated confidence bands of the $9881$ preselected genes. The darker blue lines correspond to the genes for which the estimated period of significant change in expression was determined to be between $10$ and $25$ weeks. }
    \label{full GO}
\end{figure}

Following the exclusion of $22$ genes due to measurement issues, the total number of identified genes was $9881$. For all the aforementioned genes, we assumed that the 4pLL model was the most appropriate for the analysis and applied our method to identify the time periods of significant change in gene expression.
Figure  \ref{full GO} depicts the estimated confidence bands.
The results were then employed to identify the genes with an estimated time period of significant change in expression between weeks $10$ and $25$. A total of $569$ genes were identified.

A GO enrichment analysis was conducted on the aforementioned genes using the topGo \texttt{R} package (see \cite{AR24} and  \cite{AR06}). 
In essence, Fisher's exact test is employed to verify whether a number of interesting genes is annotated to a specific functional group that exceeds what would be expected by chance.
The $20$ gene ontology groups with the smallest adjusted p-value are presented in Table B in the supplementary material.
For instance, the majority of the genes ($525$) were assigned to the GO group 'inflammatory response'. This is in accordance with the findings that inflammation is a pivotal event in the progression of NAFLD (see \cite{WDM21} and \cite{SA20}). In \cite{WDM21} few inflammatory foci were observed in week $3$ and the number strongly increased after week $36$. Lipogranulomas, single or multiple fat globules surrounded by chronic inflammatory cells and Kupffer cells (see \cite{Bru02}), were observed as early as week $6$ and increased at week $18$.


	\section{Discussion and conclusion} \label{concl}
In this paper, we proposed a parametric method to estimate the time frame during which the expression of a gene changes significantly. Thereby, we have provided a tool, that allows for an interpretation  that is not limited to the onset of an effect or the observed time points and produces results based on a hypothesis test to the $\alpha$-level. 
As demonstrated by the 4pLL model and the beta model, the approach is applicable to a wide range of potential models, with the singular restriction that the first derivative can be derived analytically from the model function. It can be employed to analyse the expression of a single gene or, when used in conjunction with a GO enrichment analysis, to infer biological function.
Moreover, the methodology can be applied to other forms of time-response or dose-response data. By extending the method to a three-step bootstrap approach, we additionally provided the means to estimate confidence intervals for  specific time points of interest. This provides further evidence for the results of the statistical analysis and enables a comparison of multiple gene curves. 

Of course, there are limits to the method. The simulation results discussed in Section \ref{sec:ressim} show a clear tendency to underestimate the time frame of significant change in expression. The effect is relatively modest for medium levels of data variability. However, it becomes more pronounced with greater variability. 
Overall, the method would benefit from a built-in stabilization procedure against the predictable effects of a large standard deviation. 
Moreover, a potential solution for a real data set with considerable variability could be the replacement of the model selection phase with model averaging (see \cite{CH08} and \cite{Fle19}). It has been demonstrated that model averaging offers advantages over model selection, as it circumvents the potential bias introduced by model misspecifications (see \cite{Bor15}, \cite{Bre96} and, recently, \cite{HM24}).


Often, a statistical conclusion on more than one gene expression curve is of interest. In Section \ref{casestudy:GO}, we presented a possible application to a large number of genes in combination with a GO enrichment analysis. Nevertheless, a multitude of potential 
 biological questions may necessitate alternative approaches. This frequently gives rise to to a multiple testing problem. Hence, our proposed method would highly benefit from the development of a method, that corrects for this issue beyond a simple correction of the p-values.

In conclusion, we posit that our proposed method is a valuable addition to the existing tools for the analysis of gene expression data. It allows for an comprehensive analysis over the entire study period and is applicable to a wide range of expression curves.

 \section*{Supplementary material}

 Supplementary material providing additional tables and figures can be found online. Corresponding R code, which can be used to reproduce the analysis of the case study and the simulation results, is available at
 \url{https://github.com/LuciaAmeis/Identification-of-changes-in-gene-expression}.

 \section*{Funding}

This work has been supported by the Research Training Group ''Biostatistical Methods for High-Dimensional Data in Toxicology'' (RTG 2624, P7) funded by the Deutsche Forschungsgemeinschaft (DFG, German Research Foundation - Project Number 427806116).
	

\section*{Appendix}


\begin{algorithm}\label{algcit} Let $\hat{\theta}$, $\hat{\sigma}^2$, $\tilde S$ and the time points of interest be calculated from the original data set as described in Section \ref{subsec:model}. 
    \begin{itemize}
        \item[~]\textbf{Step 1}  Generate bootstrap data at the $m$ different time points using $\hat{\theta}$ and $\hat{\sigma}^2$. Do so by drawing iid. errors $\epsilon_{p,q}^*\sim N(0,1)$ and applying the formula $$y_{p,q}^*=f(t_p,\hat{\theta})+\hat{\sigma}\epsilon_{p,q}^* \text{ for } p=1,\dots,m, \ q=1,\dots,n_p.$$ Repeat this step $B_3$ times. This returns $y_{p,q}^{*,1},\dots,y_{p,q}^{*,B_3}$. 
        
        Next, calculate the estimates $\hat{\theta}_k^*$ and $\hat{\sigma}_k^{2,*}$, $k=1,\dots,B_3$, using the OLS methodology as descried above for each of the generated data sets. This yields $|f'(t,\hat\theta_k^*)|$.\\
        
    \item[~]Now for $k=1,\dots,B_3$ perform the two-step bootstrap procedure to estimate the confidence band:
\item[~]\textbf{Step 2} For $k=1,\dots,B_3$: Generate bootstrap data at the $m$ different time points using $\hat{\theta}^*_k$ and $\hat{\sigma}_k^{2,*}$. Do so by drawing   iid. errors $\epsilon_{p,q}^{**}\sim N(0,1)$ and applying the formula $$y_{p,q}^{**}=f(t_p,\hat{\theta}^*_k)+\hat{\sigma}_k^{*}\epsilon_{p,q}^* \text{for } p=1,\dots,m, \ q=1,\dots,n_p.$$ Repeat this step $B_1$ times. This returns $y_{{p,q}}^{**,1},\dots,y_{{p,q}}^{**,{B_1}}$. 

For each combination of $k=1,\dots,B_3$ and $l=1,\dots,B_1$ estimate $\hat{\theta}^{**}_{k,l}$  and $\hat{\sigma}^{2,**}_{k,l}$. This yields $|f'(t,\hat{\theta}^{**}_{k,l})|$. 

The replicates $|f'(t,\hat{\theta}^{**}_{k,1})|,\dots,|f'(t,\hat{\theta}^{**}_{{k,B_1}})|$ can be interpreted as the second level of the bootstrap. The standard error of this sample yield the estimate $\hat{\sigma}_{|f'(t,\hat{\theta}^*_k)|}$.

\item[~] \textbf{Step 3} For each combination of $k=1,\dots,B_3$ and $l=1,\dots,B_1$: Generate bootstrap data at the $m$ different time points using $\hat{\theta}^{**}_{k,l}$ and $\hat{\sigma}_{k,l}^{2,**}$. Do so by drawing   iid. errors $\epsilon_{p,q}^{***}\sim N(0,1)$ and applying the formula $$y_{p,q}^{***}=f(t_p,\hat{\theta}^{**}_{k,l})+\hat{\sigma}_{k,l}^{**}\epsilon_{p,q}^{***} \text{for } p=1,\dots,m, \ q=1,\dots,n_p.$$ Repeat this step $B_2$ times. 

Estimate $\hat \theta^{***}$ (indices omitted). This yields $B_2$ replicates of the form $|f'(t,\hat{\theta}^{***})|$. The standard errors of the $B_2$ replicates yield estimates $\hat{\sigma}_{|f'(t,\hat{\theta}^{**}_{k,l})|}$.
 \item[~]\textbf{Step 4}  For each combination of $k=1,\dots,B_3$ and $l=1,\dots,B_1$: 
 Calculate
 $$D^{*,k,l}:=\max_{t\in T}\frac{|f'(t,\hat{\theta}_{k,l}^{**})|-|f'(t,\hat{\theta^*_k})|}{\hat{\sigma}_{|f'(t,\hat{\theta}_{k,l}^{**})|}}.$$
 The respective empiric $(1-\alpha)$-quantiles of the distribution yield estimates of the critical values $c_k^*$. 
 \item[~] \textbf{Step 5} For $k=1,\dots,B_3$: Calculate the lower simultaneous confidence band
$$L^\alpha(t,\hat{\theta}^*_k)=|f'(t,\hat{\theta}_k^*)|+c_k^*\hat{\sigma}_{|f'(t,\hat{\theta}^*_k)|}.$$ 

Calculate the time frames of significant change in gene expression:
$$\tilde{S}_k\coloneqq\{t|L^\alpha(t,\hat{\theta}^*_k)>\lambda\}$$
Partition the $\tilde{S}_k$ into coherent subsets:
$$\tilde{S}_k=(S_{k,1},\dots,S_{k,s})$$
\item[~] \textbf{Step 6} For $k=1,\dots,B_3$:
Dismiss the cases with more or less coherent subsets then the original data set. Identify the time points of interest for the remaining runs. 
 Now, calculate the empiric $\frac{\alpha}{2}$- and $(1-\frac{\alpha}{2})$-quantiles  ($Q_\frac{\alpha}{2}$ and $Q_{(1-\frac{\alpha}{2})}$) for each time point of interest, respectively.  The resulting confidence intervals are:
 $$\left[Q_\frac{\alpha}{2},Q_{(1-\frac{\alpha}{2})}\right]$$ 
    \end{itemize}
\end{algorithm}

\end{document}